\documentclass[journal,twocolumn]{IEEEtran}
\usepackage{epsfig,makeidx,color}
\usepackage{amsmath,amssymb,amsthm,bbm}
\usepackage{cite,graphicx}
\usepackage{enumerate}
\usepackage{hyperref}
\hypersetup{
	colorlinks = true,
	citecolor=blue,
}

\def\cL{{\cal L}}

\def\cX{{\cal X}}

\def\rH{{\rm H}}

\def\rT{{\rm T}}

\def\uR{{\mathbb R}}
\def\uC{{\mathbb C}}
\def\uE{{\mathbb E}}

\def\uP{{\mathbb P}}
\DeclareMathOperator*{\argmin}{\arg\!\min}
\DeclareMathOperator*{\argmax}{\arg\!\max}

\newtheorem{mylemma}{\bf Lemma} 

\def\be{ \begin{equation} }
\def\ee{ \end{equation} }
\def\bea{ \begin{eqnarray} }
\def\eea{ \end{eqnarray} }

\def\by{{\bf y}}

\def\bb{{\bf b}}

\def\bg{{\bf g}}

\def\bs{{\bf s}}
\def\ba{{\bf a}}

\def\bu{{\bf u}}

\def\bn{{\bf n}}

\def\bee{{\bf e}}

\def\bA{{\bf A}}
\def\bB{{\bf B}}

\def\bG{{\bf G}}

\def\bI{{\bf I}}

\def\bR{{\bf R}}

\def\bX{{\bf X}}

\def\b0{{\bf 0}}

\def\bDelta{{\bf \Delta}}

\def\cB{{\cal B}}
\def\cC{{\cal C}}

\def\cI{{\cal I}}
\def\cN{{\cal N}}

\ifCLASSOPTIONonecolumn
  \newcommand{\figwidth}{0.60\columnwidth}
\else
  \newcommand{\figwidth}{0.95\columnwidth}
\fi

\begin{document}

\title{NOMA-based Compressive Random Access using Gaussian Spreading}

\author{Jinho Choi\\
\thanks{The author is with
the School of Information Technology,
Deakin University, Burwood, VIC 3125, Australia
(e-mail: jinho.choi@deakin.edu.au)}}

\date{today}
\maketitle

\begin{abstract}
Compressive random access (CRA) is a random access
scheme that has been considered for massive
machine-type communication (MTC) with
non-orthogonal spreading sequences, 
where the notion of compressive sensing (CS) is used 
for low-complexity detectors by exploiting the sparse
device activity. 
In this paper, we study the application
of power-domain non-orthogonal multiple access
(NOMA) to CRA in order to 
improve the performance of CRA (or
increase the number of devices to be supported).
We consider Gaussian spreading sequences and
derive design criteria through
a large-system analysis to determine the power levels for power-domain
NOMA. From simulation results, we can confirm that
the number of incorrectly detected
device activity can be reduced by applying NOMA to CRA.
\end{abstract}

{\IEEEkeywords
Gaussian spreading; Multiuser Detection; Random Access;
Machine-Type Communications (MTC)}

\ifCLASSOPTIONonecolumn
\baselineskip 26pt
\fi

\section{Introduction} \label{S:Intro}

Machine-type communication (MTC) 
becomes a key element for the Internet of Things (IoT)
as it is to support massive IoT connectivity 
in 5th generation 
(5G) networks \cite{Condoluci15} \cite{Shar15} \cite{Bockelmann16}.
There are also existing standards for MTC in Long-Term Evolution
(LTE) networks \cite{Taleb12} \cite{3GPP_MTC} \cite{3GPP_NBIoT}.
Although there are a large number of devices, 
since only a few are active at a time,
it is preferable to use an
uncoordinated transmission scheme such as
a random access scheme.
Thus, most MTC schemes are based on random access. In particular, 
ALOHA \cite{Abramson70} is mainly considered 
for MTC \cite{Arouk14} \cite{Lin14} \cite{Chang15} \cite{Choi16}.

In order to exploit the sparse activity of devices in massive MTC,
the notion of compressive sensing (CS) \cite{Donoho06}
\cite{Candes06} can be exploited 
for random access, where the 
resulting scheme is referred to as compressive random access
(CRA) \cite{Wunder14}.
CRA can also be seen as the application of code division
multiple access (CDMA) to random access, where the 
user activity detection and data detection
are carried out using CS algorithms \cite{Zhu11}
\cite{Applebaum12} \cite{Schepker13}.
The main advantage of CRA over conventional multichannel random
access schemes (e.g., multichannel ALOHA) is 
that it uses non-orthogonal
spreading sequences to effectively increase the number
of multiple access channels, while low-complexity CS-based detectors
can be used for the user activity detection \cite{Choi_CRA18}.

Existing CRA schemes can be divided into two groups.
One group is based on handshaking process 
(e.g., the random access channel (RACH) procedure in \cite{3GPP_MTC})
and the other group is based
on grant-free transmission.
When CRA is used in handshaking process,
a CS algorithm is used to detect the preambles that are
transmitted by active devices. In grant-free transmission,
each user is to have a unique spreading code and 
CRA is used to detect the signals from active users.
For example, in \cite{Beyene17},
CRA is used for the RACH procedure 
where joint channel estimation and preamble detection is carried out.
In \cite{Wunder14} \cite{Schepker15} \cite{Choi17IoT}
\cite{Wunder18} \cite{Liu18}, CRA is used for grant-free
transmission. In particular, in \cite{Wunder14} \cite{Schepker15}
\cite{Choi17IoT}, 
joint channel estimation
and data detection for CRA under a frequency-selective
fading environment is considered.
In \cite{Wunder18} \cite{Liu18}, advanced CS algorithms
are considered for the signal detection.
Furthermore,
in \cite{Choi_18Feb}, in order to support a large number of devices
with grant-free CRA, coded
sparse identification vectors are studied in conjunction with joint
data detection and device identification.

As explained in \cite{Liu18}, 
grant-free CRA can take advantage of multiple measurement vector (MMV)
setup \cite{Chen06} \cite{Davies12}, which also allows to 
derive performance limits of CRA as in \cite{Choi_CRA18},
while its performance mainly depends on the length
of spreading codes, the number of devices, the sparsity, and
the background noise.
In general, the ratio of the length of spreading codes
to the number of devices is a key parameter that
determines the number of the active devices that 
can be successfully detected.
Thus, in order to improve the performance,
a longer length of spreading codes
or a wider system bandwidth is required.

To improve the spectral efficiency,
non-orthogonal multiple access (NOMA) has been extensively studied
for cellular systems \cite{Choi14}  \cite{Ding14} \cite{Higuchi15}
\cite{Ding_CM}.
In particular, power-domain NOMA with
successive interference cancellation (SIC) is considered
to support multiple users with the same radio resource block.
The notion of NOMA can be employed for random access as in \cite{Liang17}
\cite{Choi_JSAC}, which results in a higher throughput.

In this paper, we apply power-domain NOMA to CRA 
in order to improve the performance of CRA
for a given length of spreading codes or 
a system bandwidth with multiple layers in the power domain.
This approach differs from that in
\cite{Choi_JSAC} as the spreading codes in this paper are not orthogonal.
While the application of NOMA to CRA is straightforward,
it is necessary to carefully determine 
the power levels for successful SIC with a high probability.
To this end, we consider Gaussian spreading 
(i.e., Gaussian vectors for spreading codes), which results
in a Gaussian-like interference due to the signals in the other layers.
With Gaussian spreading, we can derive design criteria 
through a large-system analysis to
determine the power levels for successful
SIC through multiple layers as in the power-domain NOMA
when the maximum a posteriori probability (MAP) detection is employed.
Provided that SIC can be successfully carried out,
the throughput of CRA can be improved by a factor of the number of
layers. However, due to the transmit power limitation
and the error propagation (in SIC), the number of layers
might be limited.

The main contributions of the paper are as follows:
\emph{i)} 
a novel CRA scheme based on NOMA is proposed to improve the performance;
\emph{ii)} design criteria are derived 
with Gaussian spreading
for successful SIC with a high probability;
\emph{iii)} a low-complexity detection method 
(to find an approximate solution to the MAP detection)
is proposed for the device activity detection.


The rest of the paper is organized as follows.
In Section~\ref{S:SM}, we present the system model
for CRA. We apply NOMA to CRA in Section~\ref{S:LCRA}
and discuss SIC.
Design criteria are derived through a large-system
analysis using Gaussian approximation in Section~\ref{S:DC}.
To find an approximate solution to the MAP
detection problem with low-complexity,
we derive an approach based on variational inference in
Section~\ref{S:VI}. 
In Section~\ref{S:Sim}, simulation results are presented.
We conclude the paper with some remarks in Section~\ref{S:Conc}.

{\it Notation}:
Matrices and vectors are denoted by upper- and lower-case
boldface letters, respectively.
The superscripts $\rT$ and $\rH$
denote the transpose and complex conjugate, respectively.
The $p$-norm of a vector $\ba$ is denoted by $|| \ba ||_p$
(If $p = 2$, the norm is denoted by $||\ba||$ without
the subscript).
For a vector $\ba$, ${\rm diag}(\ba)$
is the diagonal matrix with the diagonal elements
from $\ba$.  For a matrix $\bX$
(a vector $\ba$), $[\bX]_n$ ($[\ba]_n$) represents
the $n$th column (element, resp.).
$\uE[\cdot]$ and ${\rm Var}(\cdot)$
denote the statistical expectation and variance, respectively.
$\cC \cN(\ba, \bR)$
($\cN(\ba, \bR)$)
represents the distribution of
circularly symmetric complex Gaussian (CSCG)
(resp., real-valued Gaussian)
random vectors with mean vector $\ba$ and
covariance matrix $\bR$.

\section{System Model} \label{S:SM}

In this section, we present the system model for grant-free CRA
to support MTC.

Suppose that there are $K$ devices and a base station (BS)
in a cell. We assume that if device $k$ has a packet at a slot,
it becomes active and transmits the packet during the slot. Denote by
$x_{k,(t)} \in \cX$ the $t$th symbol of
the data packet transmitted from device $k$ to the BS,
where $\cX$ is the signal constellation. 
For convenience, we assume that the length of slot is equivalent to
the length of packet and a slot consists of $T$ symbol intervals.
Like CDMA \cite{ChoiJBook} \cite{VerduBook}, 
$\bg_k x_{k,(t)}$ is transmitted during 
a symbol interval, where $\bg_k = [g_{1,k} \ \ldots \ g_{N,k}]^\rT$
is the spreading or signature code of device $k$.
Here, $N$ is referred to as the spreading gain.
In general, the system bandwidth is proportional to $N$
when the time duration of slot is fixed.
At the BS, the received signal is given by
\be
\by_{(t)} = \sum_{k=1}^K h_k \sqrt{P_k} \bg_k x_{k, (t)} b_k + \bn_{(t)},
\ t = 1, \ldots, T,
	\label{EQ:by}
\ee
where $P_k$ is the transmit power of 
device $k$, $h_k$ is the channel coefficient
from device $k$ to the BS, and $\bn_{(t)} \sim \cC \cN(\b0, N_0 \bI)$
is the background noise. Here, $b_k
\in \{1,0\}$ is the activity variable, which becomes
1 if device $k$ becomes active and $0$ otherwise.
In addition, we assume that 
$\uE[x_{k, (t)}] = 0$ and $\uE[|x_{k, (t)}|^2] = 1$ for all $k$.
Let
$$
\bs_{(t)} = [(h_1 \sqrt{P_1} x_{1,(t)} b_1) \ \ldots 
\ (h_K \sqrt{P_K} x_{K, (t)} b_K)]^\rT.
$$
Then, $\bs_{(t)}$ is a $B$-sparse vector where $B = \sum_{k=1}^K b_k$.
The received signal can be re-written as
\be
\by_{(t)} = \bG \bs_{(t)} + \bn_{(t)},
\ t = 1, \ldots, T,
\ee
where $\bG = [\bg_1 \ \ldots \ \bg_K] \in \uC^{N \times K}$.
In general, in order to support a number of devices,
it is expected that $K \gg N$,
which means conventional multiuser detectors
(e.g., decorrelator detector) \cite{VerduBook}
cannot be used to directly estimate $\bs_{(t)}$.
However, since $\bs_{(t)}$ is sparse provided that only a fraction of
devices are being active at a time,
the signal detection 
can be carried out by the device activity detection 
as in \cite{Zhu11, Applebaum12} based on the notion of CS \cite{Donoho06,
Candes06}.
That is, the support of $\bs_{(t)}$ can be estimated,
which provides the index set of active devices.
If $B \le N$,
any multiuser detector \cite{VerduBook} can be used
with known index set of active devices\footnote{If the index set
of active devices is known, we can have a submatrix of $\bG$ by taking
the columns corresponding to the active devices. In this
case, the size of the submatrix of $\bG$ becomes $N \times B$
and the corresponding system becomes underloaded,
which means that a multiuser detector can be used to obtain
a reliable estimate of the signals from $B$ active devices \cite{VerduBook}.}
to jointly detect the signals from active devices.
Thus, it is important to have a reasonably good performance of
the device activity detection 
(to estimate the index set of active devices)
in grant-free CRA, and in this paper, we mainly focus on
the device activity detection.

Noting that the support of $\bs_{(t)}$
is the same for all $t \in \{1, \ldots, T\}$,
the notion of MMV \cite{Chen06} \cite{Davies12} 
can be exploited for the device activity detection 
\cite{Choi_ICC17} \cite{Choi_CRA18} \cite{Liu18}.
With a sufficiently large $T$ and a sufficiently
high receive signal-to-noise ratio (SNR), it is possible to detect 
up to $N-1$ active devices. 
However, in practice, the receive SNR 
depends on fading and can be low. In addition, it is expected
to keep the complexity
of the receiver algorithm low. As a result, 
it might be difficult to detect up to $N-1$ active devices.

In this paper, 
we apply the notion of NOMA to CRA to improve the performance
without increasing
system bandwidth or $N$.
Throughout the paper, we assume that
$\bg_k$ is a realization of CSCG random
vector, which is referred to as a Gaussian signature vector.
In particular, $[\bg_k]_n \sim 
\cC \cN \left( 0, \frac{1}{N} \right)$ is assumed.
In addition, we assume that
each device knows the channel coefficient (actually $|h_k|^2$),
which is estimated using the beacon signal from 
the BS in time division duplexing (TDD) mode thanks
to the channel reciprocity \cite{Bockelmann16} \cite{Choi_JSAC}.
Furthermore, the channel coefficients remain unchanged
for a slot duration (i.e., the coherence time is longer than
the slot duration).

\section{Layered CRA Approach with SIC}	\label{S:LCRA}

In this section, we apply NOMA to CRA 
and discuss key parameters.

\subsection{Application of NOMA to CRA}

\begin{figure}[thb]
\begin{center}
\includegraphics[width=\figwidth]{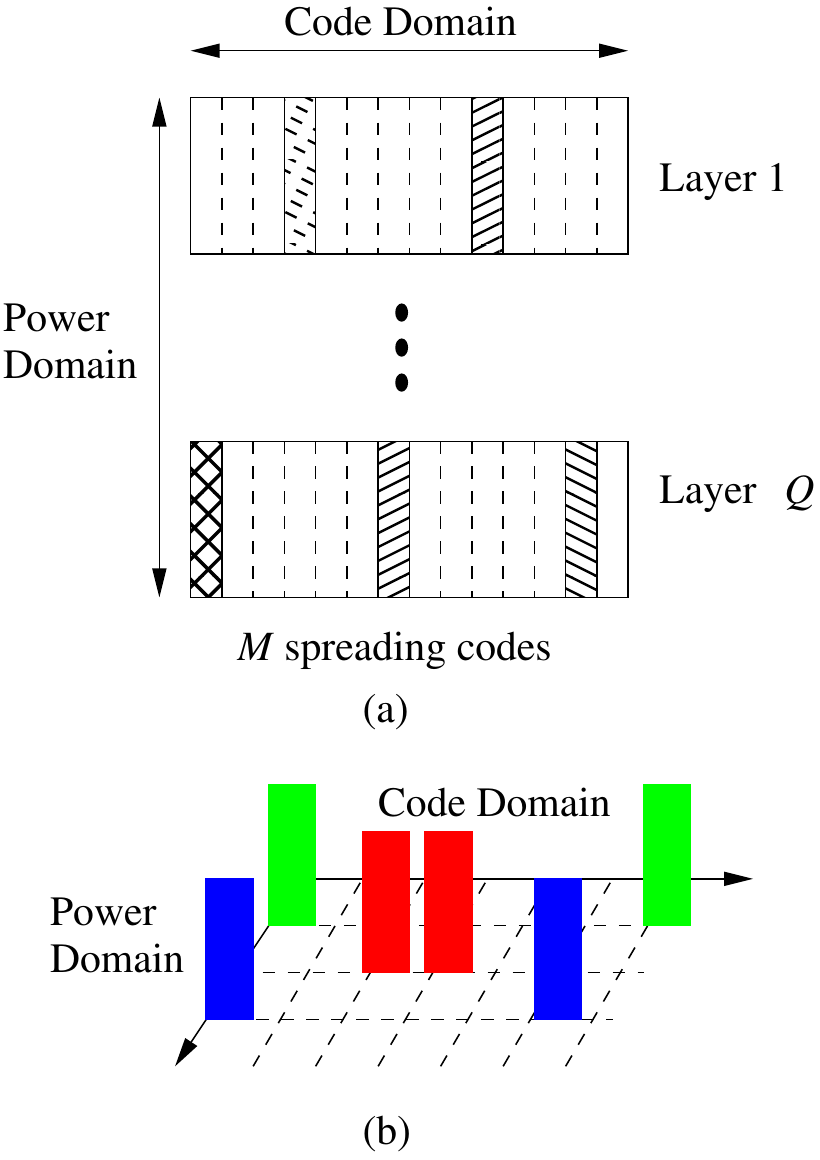}
\end{center}
\caption{The multiple access channels in the power and code domains:
(a) 2-dimensional multiple access channels; 
(b) an illustration of transmitted signals for random access in the
power and code domains.}
        \label{Fig:layers}
\end{figure}

As in \cite{Choi_JSAC},
we assume that there are 
$Q$ different layers. Each layer is characterized
by a different received signal power at the BS. 
In each layer, we assume that there are $M$ devices with unique
spreading codes 
and the resulting multiple access channels
can be characterized in the two-dimensional domain
as shown in Fig.~\ref{Fig:layers} (a). 
An illustration of transmitted signals for random access in the
power and code domains is also illustrated
in Fig.~\ref{Fig:layers} (b). 
Throughout the paper,
we assume that $K = Q M$ (i.e., in each layer, there are
$M = \frac{K}{Q}$ devices).

Suppose that an active device, say device $k$, is in the $q$th layer.
Then, the transmit power is decided to satisfy the following
received power level:
$$
P_k |h_k|^2 = V_q,
$$
where $V_q$ is the (pre-determined) received signal power for layer $q$.
Note that if $P_k$ is greater than
the maximum transmit power due to deep fading, the device
cannot be active (and needs to wait till the channel 
fading coefficient is sufficiently large).
For convenience, let
$V_1 > \ldots > V_Q \ (> 0)$.
In addition, let
$\bG_q = [\bg_{q,1} \ \ldots \ \bg_{q,M}]$,
where $\bg_{q,m}$ represents the signature code for the $m$th device
in layer $q$.
Then, the received signal is re-written as
\be
\by_{(t)} = \sum_{q=1}^Q \bG_q \bs_{q, (t)} + \bn_{(t)},
\ee
where $[\bs_{q,(t)}]_m = h_{q,m} \sqrt{P_{q,m}} x_{q,m, (t)} b_{q,m}$.
Here, $h_{q,m}$, $P_{q,m}$, $x_{q,m, (t)}$, and $b_{q,m}$
are the channel coefficient, transmit power, data symbol,
and activity variable of the $m$th device in layer $q$, 
respectively.
The resulting approach is referred to as layered NOMA-based random
access \cite{Choi_JSAC, Choi18b}.
Note that the multiple access channels in layered NOMA-based
random access in 
\cite{Choi_JSAC, Choi18b} are orthogonal. However, in this paper,
they are not orthogonal, because the spreading
codes in each layer, i.e., $\bg_{q,1}, \ldots, \bg_{q,M}$,
are not orthogonal to each other and $M > N$.
Thus, in order to differentiate the resulting scheme from 
that in \cite{Choi_JSAC, Choi18b}, we call 
it layered CRA (L-CRA).

Note that due to non-orthogonal
spreading codes with $M > N$,
the determination of power levels is not straightforward
and differs from that in \cite{Choi_JSAC}.
We will discuss the 
determination of power levels in Section~\ref{S:DC}.

\subsection{SIC and Limits of Recovery of Sparse Signals}

In this subsection, we discuss SIC to decode the signals in each 
layer successively.

Let
\begin{align}
\by_{q, (t)} 
= \by_{(t)} - \sum_{l=1}^{q-1} \bG_l \bs_{l, (t)} 
= \bG_q \bs_{q, (t)} + \bu_{q, (t)},
	\label{EQ:yq}
\end{align}
where
$$
\bu_{q, (t)} = \sum_{l=q+1}^Q \bG_l \bs_{l, (t)} + \bn_{(t)}.
$$
Since $\uE[x_{q,m, (t)}] = 0$,
we have $\uE[\bu_{q,(t)}] = 0$. 
For convenience, let $\uE[ \bu_{q, (t)}\bu_{q, (t)}^\rH] = \sigma_q^2 \bI$,
where the variance of the interference-plus-noise
vector at layer $q$, $\sigma_q^2$, is to be discussed later.
For convenience, define the SNR at layer $q$ as
$$
{\rm SNR}_q = \frac{V_q}{\sigma_q^2}.
$$
In addition,
let $B_q$ and $\rho_q$ be the number of active devices in layer $q$ and
the probability that a device in layer $q$ becomes active, i.e.,
the access probability, respectively. Clearly,
$\uE[B_q] = M \rho_q$ as there are $M$ devices in each layer.
More importantly, we can see that $\sigma_q^2$ depends on
$V_{q+1}, \ldots, V_Q$ as well as $\rho_{q+1}, \ldots, \rho_Q$.

In L-CRA, since there always exists the interference
due to the signals in higher layers, 
the success of SIC depends on the background noise and interference
(which also depends on the number of layers and power allocation).
In general, for successful SIC for all layers with a high probability,
it is necessary to carefully decide 
the power levels, $\{V_q\}$.

\section{Design Criteria for Successful SIC with Gaussian Spreading}
\label{S:DC}

In L-CRA, as a power-domain NOMA scheme,
the determination of power levels, $\{V_q\}$,
is important to guarantee successful SIC with a high probability
(which is required to avoid error propagation).
In this section, we study design criteria for the power levels 
(when an optimal device activity detection approach is employed)
through a large-system analysis under Gaussian approximations.

\subsection{Gaussian Approximation for Interference}
\label{SS:GAI}

The interference vector in \eqref{EQ:yq}, i.e., $\bu_{q,(t)}$,
is Gaussian if $q = Q$ since $\bu_{Q,(t)}$ only includes the background
noise. For $q < Q$, since Gaussian spreading
is considered (i.e.,the signature vectors,
$\{\bg_{q,m}\}$, are Gaussian),
$\bu_{q,(t)}$ can be seen as a superposition of Gaussian vectors.
Thus, for tractable analysis, we can consider 
the following assumption.
\begin{itemize}
\item[{\bf A0)}] $\bu_{q,(t)}$ is a Gaussian random vector, i.e.,
\be
\bu_{q,(t)} \sim \cC \cN(\b0, \sigma_q^2 \bI),
	\label{EQ:uq}
\ee
where
\be
\sigma_q^2 = \sum_{l=q+1}^Q V_l M \rho_l + N_0.
	\label{EQ:s2q}
\ee
\end{itemize}
In \eqref{EQ:s2q}, $V_q M \rho_q$ is the variance of the
signals in layer $q$. 

In fact, the assumption
of {\bf A0} is an approximation, because
the number of Gaussian signature vectors in each layer, i.e., 
$B_q$, is random.
In other words, the sum of a random number of independent
Gaussian random variables is not Gaussian as shown below.

\begin{mylemma}	\label{L:1}
Consider the sum of a random number of independent zero-mean
Gaussian random variables as follows:
$$
Y = \sum_{k=1}^B X_k,
$$
where $X_k \sim \cN(0, \sigma^2)$ and $B$ is a 
binomial random variable with parameters $M$ and $\rho$.
Let $Z \sim \cN(0, M\rho \sigma^2)$ be a Gaussian random variable.
Then, $Y$ is subgaussian and, for positive integer $k$,
$$
\uE[Y^k] \ge \uE[Z^k], \ \mbox{for even $k > 2$}
$$
and $\uE[Y^k] = \uE[Z^k] = 0$, for odd $k$, while 
$\uE[Y^2] =\uE[Z^2]= M \rho \sigma^2$.
\end{mylemma}
\begin{IEEEproof}
The moment generating function (MGF) of $Y$ can be found as
$$
\uE[e^{s Y}] = (1 + \rho(\phi (s) - 1) )^M,
$$
where $\phi(s) = \uE[e^{s X_k}] = e^{ \frac{\sigma^2 s^2}{2}}$.
Since $\phi(s) - 1 \ge 0$ for $s \in \uR$, we can see that
$(1 + \rho(\phi (s) - 1) )^M$ increases with $\rho$. Thus,
$$
(1 + \rho (\phi (s) - 1) )^M \le \phi(s)^M = e^{ \frac{M\sigma^2 s^2}{2}},
$$
which shows that $Y$ is subgaussian.

Let $\mu_{m,k}$ denote the $k$th moment of the zero-mean Gaussian
random variable with variance $m$. Since
\be
\mu_{m,k} = \left\{
\begin{array}{ll}
m^{k/2} (k-1)!!, 
& \mbox{if $k$ is even,} \cr
0 , & \mbox{if $k$ is odd,} \cr
\end{array}
\right.
\ee
we can show that
\begin{align}
\uE[Y^k] & = \sum_{m=0}^M p_m (M) \uE[(X_1 + \ldots + X_m)^k] \cr
& = \sum_{m=0}^M p_m (M) \sigma^{k} \mu_{m,k} \cr
& = \left\{
\begin{array}{ll}
\uE[B^\frac{k}{2}] \sigma^{k} (k-1)!! ,
& \mbox{if $k$ is even,} \cr
0 , & \mbox{if $k$ is odd.} \cr
\end{array}
\right.
	\label{EQ:momY}
\end{align}
where
$\uE[B^l] = \sum_{m=0}^M m^l p_m (M)$ and 
$p_m (M) = \binom{M}{m} \rho^m (1-\rho)^{M-m}$.

Let $Z$ be the zero-mean Gaussian random variable with
variance $M\rho \sigma^2$. 
Then, the moments of $Z$ are given by \cite{Papoulis}
\begin{align}
\uE[Z^k] 
= \left\{
\begin{array}{ll}
(M \rho)^{k/2} \sigma^{k} (k-1)!!, & \mbox{if $k$ is even}, \cr
0, & \mbox{if $k$ is odd.}\cr
\end{array}
\right.
	\label{EQ:momZ}
\end{align}
From \eqref{EQ:momY} and \eqref{EQ:momZ}, the 2nd moments of $Y$ and $Z$ are 
the same, because $\uE[B] = M \rho$.
Furthermore, using Jensen's inequality, we have
$$
\uE \left[
\left(\frac{B}{M \rho} \right)^l
\right] \ge 
\left(
\uE \left[
\frac{B}{M \rho}
\right]  \right)^l = 1, \ l \in \{0,1,\ldots\},
$$
which implies that $\uE[B^l] \ge (M \rho)^l$. Applying this to
\eqref{EQ:momY}
and \eqref{EQ:momZ}, we can see that $\uE[Y^k] \ge \uE[Z^k]$ for all 
positive even integers $k$,
which completes the proof.
\end{IEEEproof}

\begin{figure}[thb]
\begin{center}
\includegraphics[width=\figwidth]{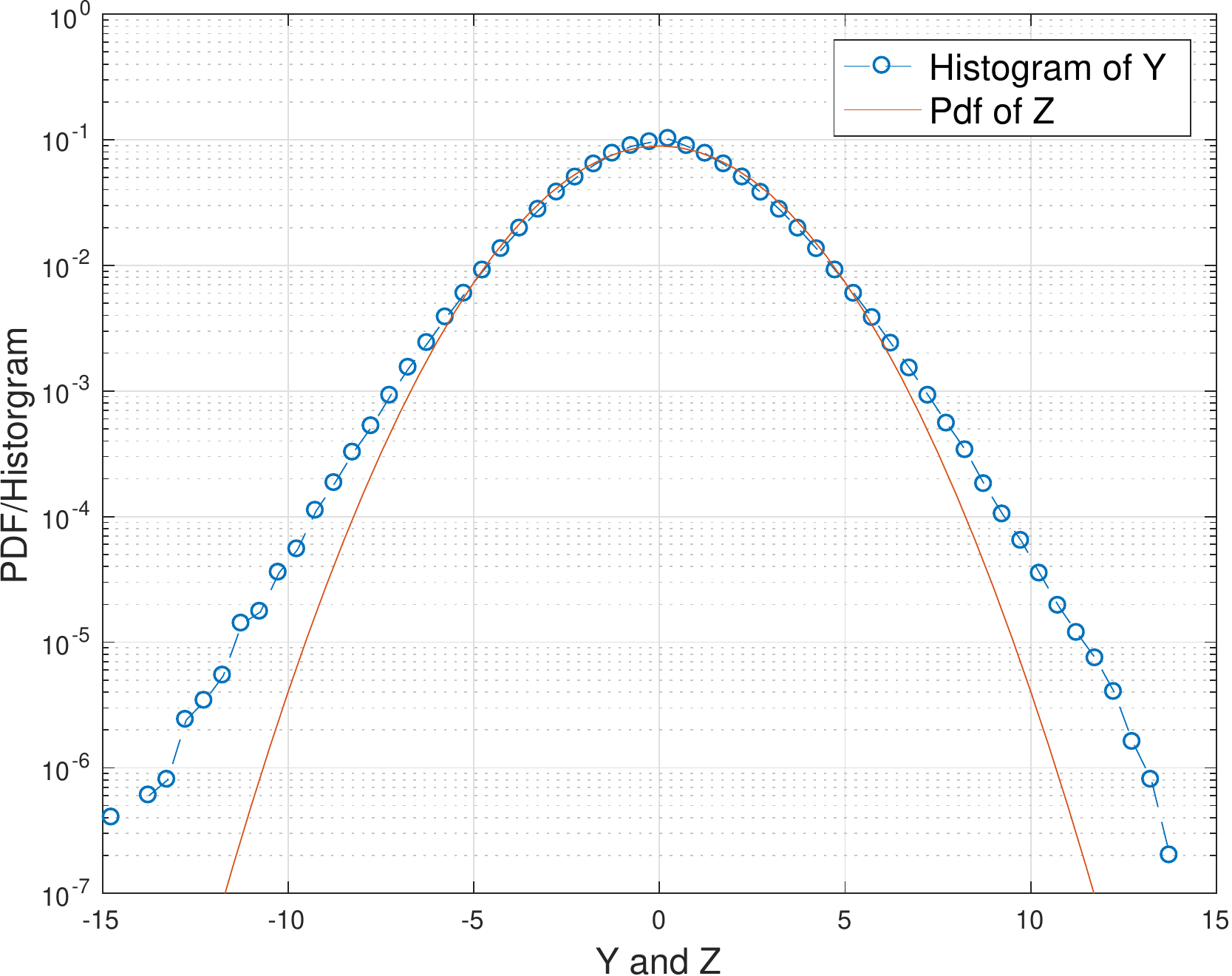}
\end{center}
\caption{The pdf and histogram of $Z$ and $Y$, respectively,
when $\sigma^2 = 1$ and $M = 100$ and $\rho = 0.05$.}
        \label{Fig:plt_RSG}
\end{figure}

According to Lemma~\ref{L:1}, 
since the moment of $Y$ is higher than or equal to 
that of $Z$, we expect that 
the tail probability of $Y$ is heavier
than that of Gaussian $Z$.
In Fig.~\ref{Fig:plt_RSG}, we show the 
probability distribution function
(pdf) of $Z$
and the histogram of $Y$ when $\sigma^2 = 1$ and $M = 100$ and 
$\rho = 0.05$.
Although the variances of $Y$ and $Z$ are
the same, we can see that the tail probability of $Y$ is heavier
than that of Gaussian $Z$.
However, since $Y$ is subgaussian, the tail probability is also an exponential
function.

In summary,
although we consider 
the Gaussian approximation for $\bG_q \bs_{q,(t)}$ 
for tractable analysis, as shown above, it is expected that 
it results in an optimistic performance prediction.

\subsection{An Optimal Approach for the Device Activity Detection under
Gaussian Approximations}	\label{SS:OA}

In this subsection, we consider an optimal approach for
the device activity detection to derive design
criteria. In particular, we consider the MAP
approach \cite{ChoiJBook2}.

Consider \eqref{EQ:yq} under the assumption that the signals
in layers $1, \ldots, q-1$ are perfectly recovered and removed.
Let 
$$
\cI_q = {\rm supp} (\bs_{q, (t)}), \ t = 1,\ldots, T.
$$
The device activity detection is to detect (or estimate) $\cI_q$
from $\by_q = [\by_{q, (1)}^\rT \ \ldots \ \by_{q, (T)}^\rT]^\rT$.
We note that
$$
\uE[|[\bs_{q, (t)}]_k|^2] = \left
\{
\begin{array}{ll}
V_q, & \mbox{if $k \in \cI_q$} \cr
0, & \mbox{o.w.} \cr
\end{array}
\right.
$$
Thus, we have the following assumption in order to derive
a tractable approach for the device activity detection.

\begin{itemize}
\item[{\bf A1)}] For $k \in \cI_q$, we assume that
$[\bs_{q, (t)}]_k$ is an independent zero-mean CSCG random variable
with variance $V_q$,
which is referred to as the Gaussian approximation \cite{Lin15} for 
the signal. 
\end{itemize}
Under the Gaussian approximations 
for signal and interference (i.e., the assumptions of {\bf A1}
and {\bf A0}, respectively),
$\by_q$ becomes a Gaussian random vector with
the following covariance matrix that depends on $\bb_q
= [b_{q,1} \ \ldots \ b_{q,M}]^\rT \in \cB$:
$$
\bR_q (\bb_q) = \bG_q \bB_q \bG_q^\rH + \sigma_q^2 \bI,
$$
where $\bB_q = {\rm diag}(\bb_q)$.
Here, $\cB = \{ \bb\,|\, [\bb]_m \in \{0,1\}\}$.
As a result, the likelihood
function of $\bb_q$ for given $\by_q$ is given by
\be
f(\by_q\, |\, \bb_q) = \frac{
\exp\left( -\sum_{t=1}^T \by_{q,(t)}^\rH 
\bR_q(\bb_q)^{-1} \by_{q,(t)}
\right)}{\pi^N \det(\bR_q(\bb_q))}.
	\label{EQ:lh}
\ee
Since the a posteriori probability is 
given by
$$
\Pr(\bb_q\,|\, \by_q) = C f(\by_q\, |\, \bb_q) \Pr(\bb_q),
$$
where $C$ is a constant,
the MAP approach 
for the device activity detection
can be given by
\begin{align}
\hat \bb_q 
& = \argmax_{\bb_q \in \cB} \Pr(\bb_q\,|\, \by_q) \cr
& = \argmax_{\bb_q \in \cB}
\ln f(\by_q\, |\, \bb_q) + \ln \Pr(\bb_q).
	\label{EQ:map}
\end{align}
Throughout the paper, we assume that
each device becomes independently active. Thus, we have
the following assumption.
\begin{itemize}
\item[{\bf A2)}] $\Pr(\bb_q)$ 
can be represented by a binomial distribution as follows:
\be
\Pr(\bb_q) = \prod_{m=1}^M \Pr(b_{q,m}) = 
\binom{M}{B_q} \rho_q^{B_q} (1- \rho_q)^{M-B_q},
\ee
where $\rho_q$ is the probability
of being active for the devices in layer $q$ and $B_q = ||\bb_q||_0$.
\end{itemize}

Since $\bb_q$ is a binary vector of length $M$,
the computational complexity 
of the MAP approach
is proportional to $|\cB|  = 2^M$,
which might be prohibitively high when there are a large
number of devices (as $M = K/Q$),
although it is to be used to derive design criteria (in 
Subsection~\ref{SS:LS}).
To avoid this problem, we will consider a low-complexity
approach in Section~\ref{S:VI}.

\subsection{A Large-System Analysis}	\label{SS:LS}

For convenience, we omit the layer index $q$.
Denote by $\bb$ the true activity vector.
Consider two vectors $\bb, \bb^\prime \in \cB$,
where $\bb \ne \bb^\prime$.
Let 
$\by = [\by_{(1)}^\rH \ \ldots \ \by_{(T)}^\rH]^\rH$
and
$$
\cL_{\rm ap} (\bb) = \cL (\bb) + \ln \Pr(\bb),
$$
where $\cL (\bb) = \ln f(\by\,|\, \bb)$.

When the MAP detection is considered,
from \eqref{EQ:map}, the pairwise error probability
(PEP) \cite{BiglieriBook} is given by
\begin{align}
P(\bb \to \bb^\prime) & = 
  \Pr( \cL_{\rm ap} (\bb) < \cL_{\rm ap} (\bb^\prime) \,|\, \bb) \cr
& = \Pr\left(\sum_t \by_{(t)}^\rH \bDelta \by_{(t)} >  
d(\bb, \bb^\prime)
 \,|\, \bb \right) ,
	\label{EQ:PEP}
\end{align}
where
$\bDelta = \bR(\bb)^{-1} - \bR(\bb^\prime)^{-1}$ and
\begin{align}
d(\bb, \bb^\prime) 
& = \ln \Pr(\bb) - \ln \Pr(\bb^\prime) \cr
& \quad - T (\ln \det(\bR(\bb))-
\ln \det(\bR(\bb^\prime))).
	\label{EQ:dxx}
\end{align}
We can have a useful representation of 
$\sum_t \by_{(t)}^\rH \bDelta \by_{(t)}$  
to find the PEP for certain cases as follows.

\begin{mylemma}	\label{L:bee}
Let $\bee = \bb^\prime - \bb$ and suppose
that ${\rm supp}(\bee) = l$ (i.e., $\bb$ differs from
$\bb^\prime$ by only one element).
Then, under the assumption of {\bf A1},
for given $\bb$, we have
\be
\sum_{t=1}^T \by_{(t)}^\rH \bDelta \by_{(t)} = 
\left\{
\begin{array}{ll}
\frac{1}{2}
\frac{V \alpha_l}{1 + V \alpha_l } \chi_{2T}^2
\ (\ge 0),
& \mbox{if $e_l = 1$;} \cr
-\frac{1}{2}
V \alpha_l^\prime \chi_{2T}^2 \ (\le 0),
& \mbox{if $e_l = -1$,} \cr
\end{array}
\right.
	\label{EQ:y_chi}
\ee
where $\chi_n^2$ represents a chi-squared random variable
with $n$ degrees of freedom and
\be
\alpha_l = \bg_l^\rH \bA (\bb) \bg_l \ \mbox{and}\
\alpha_l^\prime = \bg_l^\rH \bA(\bb^\prime)\bg_l.
	\label{EQ:alal}
\ee
Here, $\bA (\bb) = \bR(\bb)^{-1}$.
\end{mylemma}
\begin{IEEEproof} 
See the proof of Lemma 3 in \cite{Choi_MCMC}.
\end{IEEEproof}

The case of $e_l = +1$ corresponds to the event of false alarm (FA)
where device $l$ is not active, but the receiver decides that it is active,
while that of $e_l = -1$ 
corresponds to the event of missed detection (MD) where device $l$
is active, but the receiver does not see it.
Thus, in Lemma~\ref{L:bee},
we only consider the event of one MD or one FA. Although
there can be the events of more than one MDs or FAs or
mixed MDs and FAs, their probabilities
might be sufficiently low to ignore,
compared to that of the event of one MD or one FA. 

As derived in \cite{Choi_MCMC},
when ${\rm supp}(\bee) = e_l = +1$,
we can show that
\begin{align*}
d(\bb, \bb^\prime) 
& = d_{\rm FA} (||\bb||_0) \cr
& = T \ln (1+ V \alpha_l) + 
\ln \frac{||\bb||_0 +1}{M-||\bb||_0}
+ \ln \frac{1 - \rho}{\rho} . 
\end{align*}
In addition, when $e_l = -1$,
it follows
\begin{align*}
d(\bb, \bb^\prime) 
= - d_{\rm MD} (||\bb||_0),
\end{align*}
where
$$
d_{\rm MD} (||\bb||_0) = 
T \ln (1+ V \alpha_l^\prime) - \ln \frac{M-||\bb||_0 +1}{||\bb||_0}
- \ln \frac{\rho}{1 - \rho}.
$$
Thus, by substituting \eqref{EQ:y_chi} into \eqref{EQ:PEP},
the PEPs of FA and MD are given by
\begin{align}
\uP_{\rm FA} (||\bb||_0)
& = \Pr\left(\chi_{2T}^2 > \frac{2 d_{\rm FA} (||\bb||_0)}
{\xi_{l}} \right)  \cr
\uP_{\rm MD} (||\bb||_0) 
& = \Pr\left(\chi_{2T}^2 < \frac{2 d_{\rm MD} (||\bb||_0)}
{V \alpha_l^\prime } \right) , 
\end{align}
where 
$\xi_l = \frac{\alpha_l} {1+ \alpha_l}$.

We now consider a large-system 
with $N \to \infty$ \cite{Tse99}.
Since $||\bb||_0$ is the number of active devices,
as in \cite{Tse99}, we assume that 
$\frac{||\bb||_0}{N}$ approaches a constant as follows:
$$
\frac{||\bb||_0}{N} \to \kappa = \frac{\rho M}{N},
\ N \to \infty.
$$
From \eqref{EQ:alal}, it can be shown that
\be
\alpha_l = 
\gamma \bg_l^\rH 
\left(\bI + \gamma \tilde \bG \tilde \bG^\rH \right)^{-1}
\bg_l,
\ee
where $\tilde \bG$ is the submatrix of $\bG$ with the columns
corresponding to the support of $\bb$ and
$\gamma = \frac{1}{\sigma^2}$.
According to \cite{Tse99},
$\alpha_l$ can be seen as the signal-to-interference
ratio (SIR) of the minimum mean-squared error (MMSE) receiver.
When the elements of $\bg_l$ are 
independent and identically distributed (i.i.d.)
and $\bg_l$ and $\tilde \bG$ are independent,
for approximation, we can use the mean of
$\alpha_l$ (where the expectation is 
carried out over the elements of $\bg_l$ and $\tilde \bG$) that is
given by
$$
\alpha_l \approx \uE[ \alpha_l] = \beta(\gamma, ||\bb||_0/N) , 
$$
where $\beta(\cdot, \cdot)$ is derived in \cite{Tse99} as follows:
\begin{align}
\beta(\gamma, \kappa)
= \frac{(1-\kappa)\gamma}{2} - \frac{1}{2} 
+ \sqrt{
\frac{(1-\kappa)^2 \gamma^2}{4} +
\frac{(1+\kappa) \gamma}{2} + \frac{1}{4} }.
\end{align}
We can also replace 
$\alpha_l^\prime$ with its mean for approximation
to find $\uP_{\rm MD}$ as follows:
$$
\alpha_l^\prime \approx
\uE[\alpha_l^\prime] = \beta(\gamma, ||\bb^\prime||_0/N).
$$
Since $\frac{||\bb||_0}{N},
\frac{||\bb^\prime||_0}{N} \to \kappa$ as $N \to \infty$, 
we have
\be
\alpha_l, \alpha_l^\prime \to \beta(\gamma, \kappa)
\ee
in a large-system.
Furthermore, we have
\begin{align}
\ln \frac{||\bb||_0 + 1}{M - ||\bb||_0} 
& \to \ln \frac{\rho}{1 - \rho} \cr
\ln \frac{M-||\bb||_0+1}{||\bb||_0}
& \to  \ln \frac{1 - \rho}{\rho} . 
\end{align}
Consequently, in a large-system, we have
\begin{align}
\frac{d_{\rm FA} (||\bb||_0)}{T \xi_l} & \to 
\frac{(1+\beta)\ln (1+ V \beta)}{\beta} \cr
\frac{d_{\rm MD} (||\bb||_0)}{T V \alpha_l^\prime} & \to 
\frac{\ln (1+ V \beta )}{V \beta},
\end{align}
where $\beta = \beta(\gamma, \kappa)$.
The asymptotic probabilities of (one) FA and (one) MD become
\begin{align}
\tilde \uP_{\rm FA} & = 
\Pr \left( \frac{\chi_{2T}^2}{2T} > 
\frac{(1+\beta)\ln (1+ V \beta)}{\beta} \right)  \cr
\tilde \uP_{\rm MD} & = 
\Pr \left( \frac{\chi_{2T}^2}{2T} < 
\frac{\ln (1+ V \beta)}{V \beta} \right).
	\label{EQ:APs}
\end{align}
In \eqref{EQ:APs}, 
since we have $\uE \left[ \frac{\chi_{2T}^2}{2T} \right] = 1$,
low error probabilities are expected for a sufficiently large
$T$ if
\begin{align}
\frac{(1+\beta)\ln (1+ V \beta)}{\beta} > 1 \cr
\frac{\ln (1+ V \beta)}{V \beta} < 1.
	\label{EQ:cond}
\end{align}
According to \cite{Laurent00}, for example,
we can have the following bound:
$$
\tilde \uP_{\rm MD} \le
\exp \left(
- \frac{T}{2} \left(1 - 
\frac{\ln (1+ V \beta)}{V \beta} \right)
\right).
$$
Thus, if the conditions in \eqref{EQ:cond} hold, 
$\tilde \uP_{\rm MD}$ can exponentially decrease with $T$
(which is also valid   
$\tilde \uP_{\rm FA}$) in a large-system.
We can have the following result to determine $V$.

\begin{mylemma}
If 
\be
V \beta > 1 \ \mbox{and} \  V \ge 1, 
	\label{EQ:Vb}
\ee
\eqref{EQ:cond} holds.
\end{mylemma}
\begin{IEEEproof}
Since 
\be
\frac{x}{1+x} \le \ln (1+ x) \le x, \ x > -1,
	\label{EQ:lnx}
\ee
we can show that
$$
\frac{(1+\beta)\ln (1+ V \beta)}{\beta} \ge 
\frac{(1+\beta) V \beta}{\beta (1+ V\beta)} = \frac{V(1+ \beta)}{1+ V\beta}
> 1,
$$
which shows that the first inequality in \eqref{EQ:cond} holds.
To show the second inequality, we can apply the second in equality
in \eqref{EQ:lnx}, which completes the proof.
\end{IEEEproof}

\subsection{Determination of Power Levels}

In this subsection, we consider a cell and discuss an approach to 
determine the power levels 
according to the conditions in \eqref{EQ:Vb}.

\begin{figure}[thb]
\begin{center}
\includegraphics[width=\figwidth]{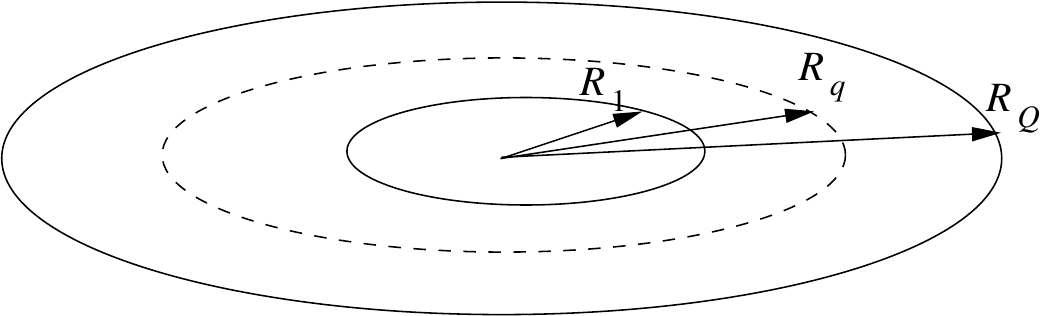}
\end{center}
\caption{An illustration of dividing a cell into $Q$ regions
to reduce the transmit power.}
        \label{Fig:cell}
\end{figure}

As shown in Fig.~\ref{Fig:cell},
suppose that a cell is divided into $Q$ regions. 
Region $q$ is a circular ring with 
the inner and outer radii, denoted by $R_{q-1}$ and $R_q$,
respectively, where $R_{0} = 0$.
Thus, the transmit power of a device in region
$q$ needs to be higher than that in region $q-1$ for
the same receive power.
Thus, in order to avoid a high transmit power,
the devices in region $q$ can be assigned to layer $q$
\cite{Choi_JSAC}.

Suppose that $K$ devices are uniformly distributed in a cell.
Then, $\{R_q\}$ can be decided
to make the area of each region equal (so that each area has the same
number of devices (on average), $M = \frac{K}{Q}$).
For normalization purposes, we assume that $R_1 = 1$.
Since the power control is considered
for the pre-determined receive power levels,
for a device in region or layer $q$, we expect
$$
P_{q,m} \propto R_q^\eta V_q,
$$
where $\eta$ is the path loss exponent.
Here, we consider a device on the outer ring 
(i.e., the worst case) and $|h_k|^2 
\propto \frac{1}{R_q^\eta}$.

With a target SNR, denoted by $\Gamma$,
for given access probabilities, $\{\rho_q\}$,
the power levels can be recursively decided as follows:
\be
V_q = \frac{ \Gamma}{ \beta \left(
\frac{1}{\sigma_q^2}, \kappa \right)}, \ q = Q, \ldots, 1.
	\label{EQ:Vq}
\ee
Here, for $q = Q$, since
$V_Q = \frac{ \Gamma}{ 
\beta \left(\frac{1}{N_0}, \kappa \right)}$,
we need to have
\be
\Gamma \ge \beta \left(\frac{1}{N_0}, \kappa \right)
\ee
to make sure that $V_Q \ge 1$.
Note that since $\beta(\gamma, \kappa)$ increases with $\gamma$
and $\sigma_q^2 > \sigma_{q+1}^2$,
from \eqref{EQ:Vq}, we can see that $V_1 > \ldots > V_Q$ as expected.

For example, suppose that $Q = 3$, 
$M = 100$, $N = 30$, and $\rho_q = 0.05$ for all $q$.
Then, there might be 15 active devices on average.
In addition, let $\Gamma = 4$, $\eta = 3.5$, and $N_0 = 1$.
In Table~\ref{TBL:1}, the pre-determined power levels, $\{V_q\}$,
and the average transmit powers for three layers
are shown, where we can see that 
by taking advantage of shorter distances
to the BS, the transmit power
of the devices in region $q$ can be lower than $V_q$ for $q < Q$.

\begin{table}[thb]
\caption{The transmit power for each region and the pre-determined
power level for each layer.}
\begin{center}
\begin{tabular}{|c||l|r|r|} \hline
Layer & $R_q$ & $V_q$ (dB) & Transmit Power (dB) \cr \hline \hline
1 & 0.577 & 32.840 & 24.490 \cr
2 & 0.816 & 19.618 & 16.536 \cr
3 & 1 & 6.382 &   6.382 \cr \hline
\end{tabular}
\end{center}
\label{TBL:1}
\end{table}

In this section, although we only consider the determination of
power levels to satisfy \eqref{EQ:cond},
it might be possible to jointly
determine the power levels and access probabilities.
We also note that the average transmit power of the devices
in layer 1 might be too high 
although the advantage of the short distance to the BS in layer 1
is taken into account
when $Q = 3$ as shown in Table~\ref{TBL:1}.
Thus, $Q$ may not be too large
to avoid a high transmit power (e.g., we may have L-CRA with
two layers in practice).

\section{Variational Inference based Low-Complexity Detection}	\label{S:VI}

For the device activity detection,
we use 
any low-complexity CS algorithm that can exploit
the sparse activity as in \cite{Applebaum12} 
\cite{Schepker15} \cite{Wunder18} \cite{Choi_CRA18} \cite{Liu18}.
However, in Subsection~\ref{SS:OA},
since the MAP approach is considered for the design
criteria,
it might be desirable to consider the MAP detector.
Unfortunately, its complexity is prohibitively high
as mentioned earlier and we need to resort to low-complexity
approximations.
In this section, we consider a variational inference
technique, namely 
the coordinate ascent variational inference (CAVI) algorithm
\cite{Bishop06, Blei17}, which 
can provide a low-complexity approximate solution to the MAP
detection problem in \eqref{EQ:map}.

For convenience, we omit the layer index $q$.
In \eqref{EQ:map}, the activity variables, which are binary 
random variables, are to be detected. 
If an exhaustive search is considered, the complexity
is proportional to $|\cB| = 2^M$ for each layer.
To avoid this, we can consider the variational
distribution for each $b_{m}$, denoted by $\psi_m(b_m)$,
and solve the following optimization problem:
\be
\psi^* (\bb) = \argmin_{\psi(\bb) \in \Psi}
{\rm KL} (\psi (\bb) || \Pr(\bb\,|\, \by)),
	\label{EQ:VI}
\ee
where $\psi(\bb) = \prod_{m} \psi(b_m)$, 
$\Psi$ represents the collection of all the possible distributions of 
$\bb$, and ${\rm KL}(\cdot)$ is
the Kullback-Leibler (KL) divergence that is defined as
$$
{\rm KL} (\psi(\bb)||f(\bb))
= \sum_\bb \psi(\bb) \ln \frac{\psi(\bb)}{f(\bb)}.
$$
Here, $f(\bb)$ is any distribution of $\bb$
with $f(\bb) > 0$ for all $\bb \in \cB$.
In \eqref{EQ:VI},
clearly, we attempt to find $\psi(\bb)$ that is close
to the a posteriori probability, $\Pr(\bb\,|\, \by)$,
as an approximation.
In \cite{Blei17}, 
the minimization of 
the KL divergence is equivalent to 
the maximization of the evidence lower bound (ELBO), which is given by
$$
{\rm ELBO}(\psi) = \uE[\ln f(\by,\bb)] - \uE[\ln \psi(\bb)].
$$
Let $\bb_{-m} = [b_1 \ \ldots \ b_{m-1} \ b_{m+1} \ \ldots \ b_M]^\rT$
and
$\psi_{-m} (\bb_{-m}) = \sum_{l \ne m} \psi_l (b_l)$.
The CAVI algorithm \cite{Bishop06, Blei17} is to 
update one variational distribution at a time 
with the other variational distributions fixed
(so that the ELBO can be minimized through iterations) as follows:
\begin{align}
\psi_m^*(b_m) 
& \propto \theta_m (b_m) \cr
& = \exp \left(\uE_{-m} [\ln f(b_m\,|\, \bb_{-m}, \by)] \right),
	\label{EQ:theta}
\end{align}
where $\uE_{-m}[\cdot]$ 
represents the expectation with 
respect to $\bb_{-m}$ or with the distribution $\psi_{-m} (\bb_{-m})$.
Let $\psi_m^{(i)}$ denote the $i$th estimate of $\psi_m$.
In the CAVI algorithm, $\psi_m^{(i)}$ is updated from $m = 1$ to $M$ in each
iteration. The number of iterations is denoted by $N_{\rm run}$.
Unfortunately, the convergence behavior of the CAVI algorithm
is not known \cite{Blei17} and $N_{\rm run}$ can be decided 
through experiments.


Thanks to the Gaussian approximations,
it is possible to find a closed-form expression
for $\theta_m (b_m)$ in \eqref{EQ:theta}, 
which can be found in \cite{Choi_VISM},
where it is also shown that the complexity of the CAVI
algorithm per iteration is $O(M N^2)$.

\section{Simulation Results}	\label{S:Sim}

In this section, we present simulation results for L-CRA.
We assume that a cell is divided into $Q$ regions and
the devices in region $q$ can perform the power control
so that the received powers at the BS become $V_q$.
In addition, the power levels are decided as in \eqref{EQ:Vq}
with $\eta = 3.5$ and $N_0 = 1$,
and in each simulation run, Gaussian spreading codes are randomly
generated.

For the device activity detection at the BS, we use the CAVI algorithm.
To see the performance, we consider the numbers of MD and FA devices,
where an MD device is an active device that is not detected
and an FA device is an inactive device that is detected
(as an active device).
Throughout this section, for convenience, we assume that the BS
knows the number of active devices in each layer, $B_q$,
and carry out the device activity detection by choosing the devices
associated with the $B_q$ largest $\psi_m(1)$'s in each layer.
In this case, the number of MD devices and that of FA devices are the same.
Note that in L-CRA with $Q > 1$, there is error propagation
due to erroneous SIC in the presence of FA as well as MA devices.
Thus, the presence of either FA or MD devices degrades the performance,
while MD devices can lead a worse performance degradation
than FA devices (in terms
of the throughput\footnote{Note that in this paper,
we do not consider the throughput as no re-transmission
strategies are considered}), because 
MD devices need to re-transmit their packets.

\begin{figure}[thb]
\begin{center}
\includegraphics[width=\figwidth]{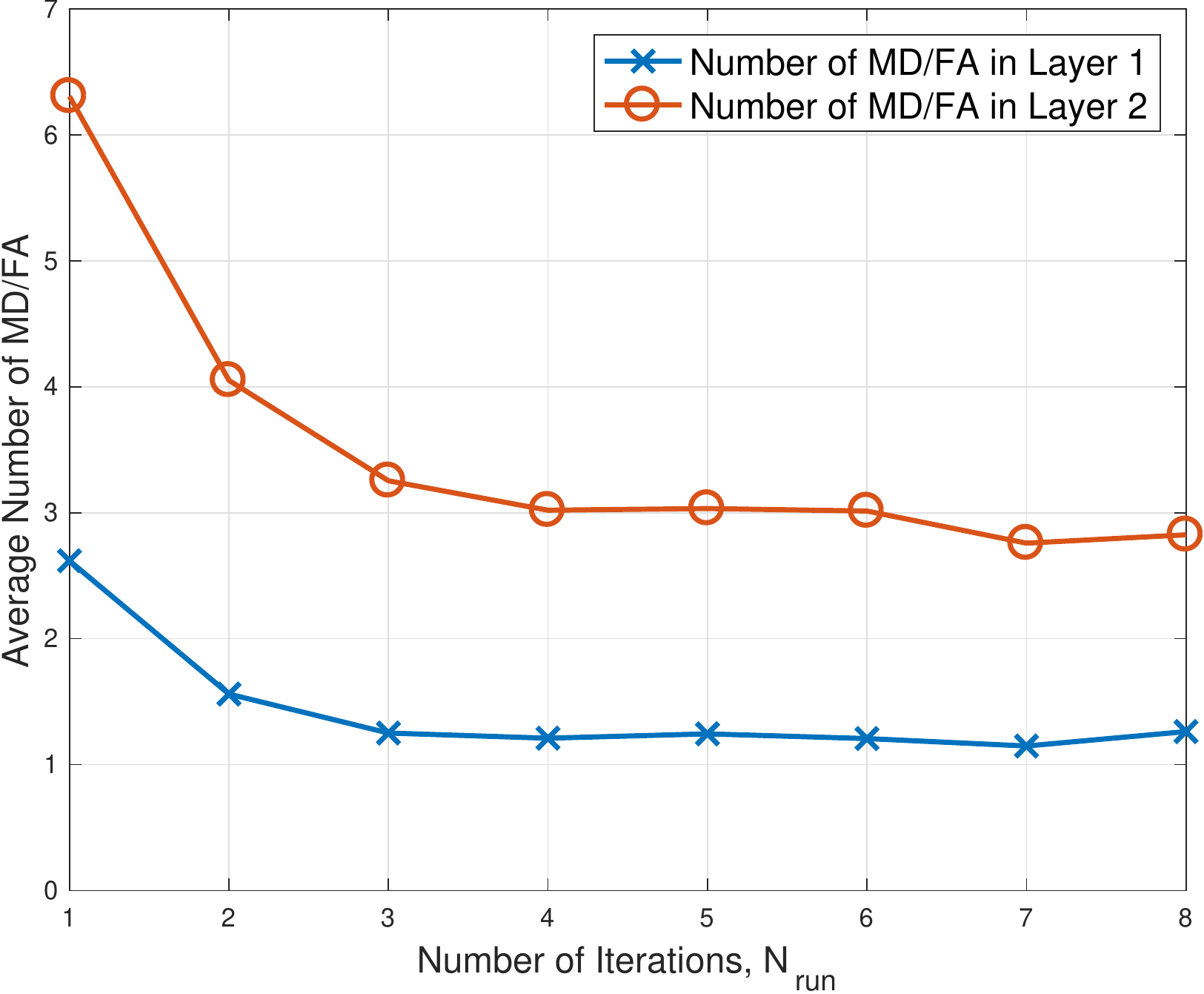} 
\end{center}
\caption{The average number of MD/FA devices
as a function of the number of iterations for CAVI, $N_{\rm run}$,
when $Q = 2$, $M = 150$, 
$N = 30$, $\rho = 0.05$, $\Gamma = 4$, and $T = 100$.} 
        \label{Fig:plt2}
\end{figure}

Before we see the impact of key system parameters
(e.g., access
probability $\rho$, target SNR $\Gamma$, and spreading gain $N$)
on the performance, we consider the number of iterations of the CAVI algorithm
to see the convergence.
In Fig.~\ref{Fig:plt2}, we show 
the average number of MD/FA devices
as a function of the number of iterations for CAVI, $N_{\rm run}$,
when $Q = 2$, $M = 150$, 
$N = 30$, $\rho = 0.05$, $\Gamma = 4$, and $T = 100$. 
Clearly, as expected,
a better performance is achieved with more 
iterations. However, we can see that $N_{\rm run} \ge 4$ might be sufficient
as the performance improvement becomes negligible with increasing
$N_{\rm run}$ once $N_{\rm run} \ge 4$.

\begin{figure}[thb]
\begin{center}
\includegraphics[width=\figwidth]{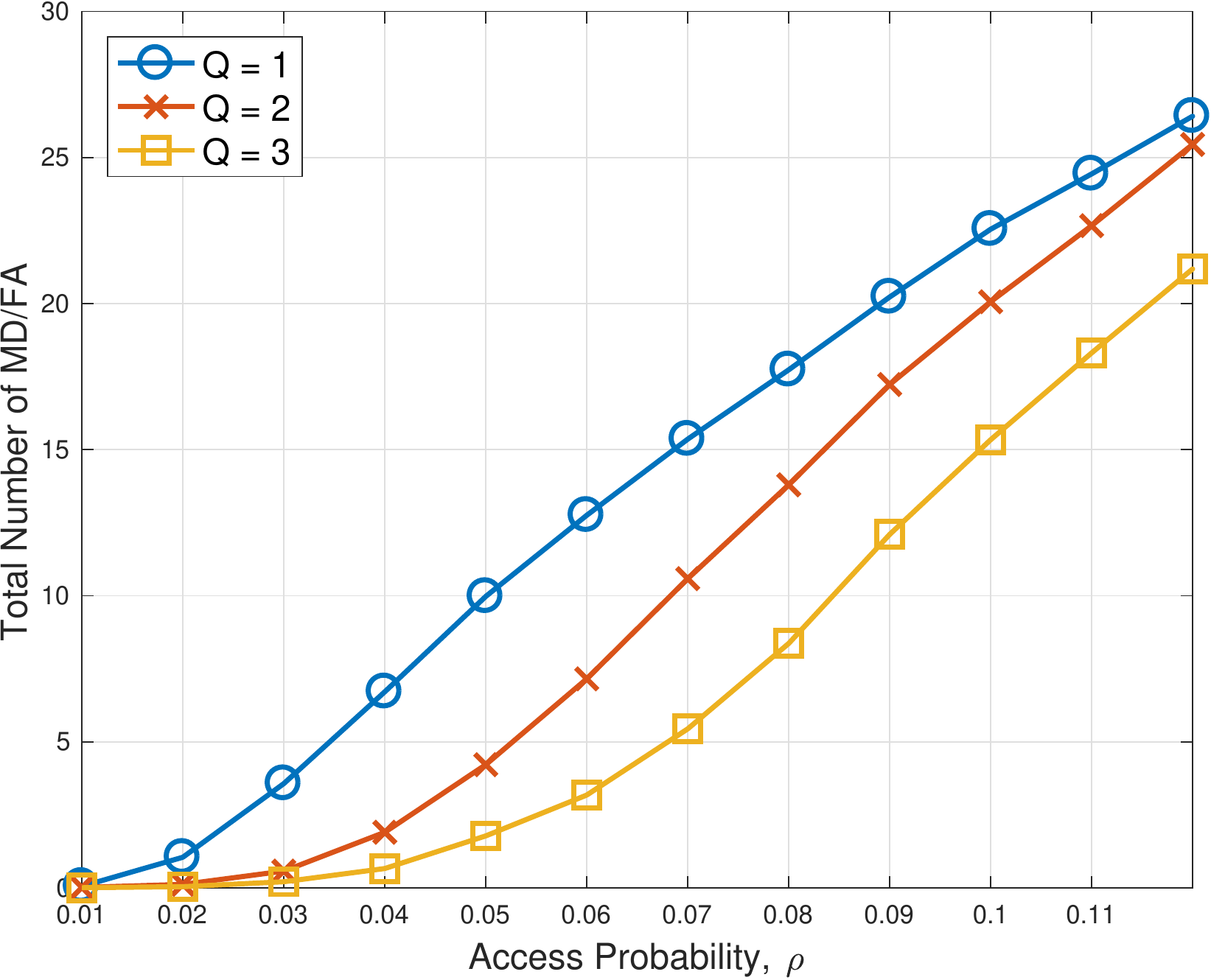} \\
(a) \\
\includegraphics[width=\figwidth]{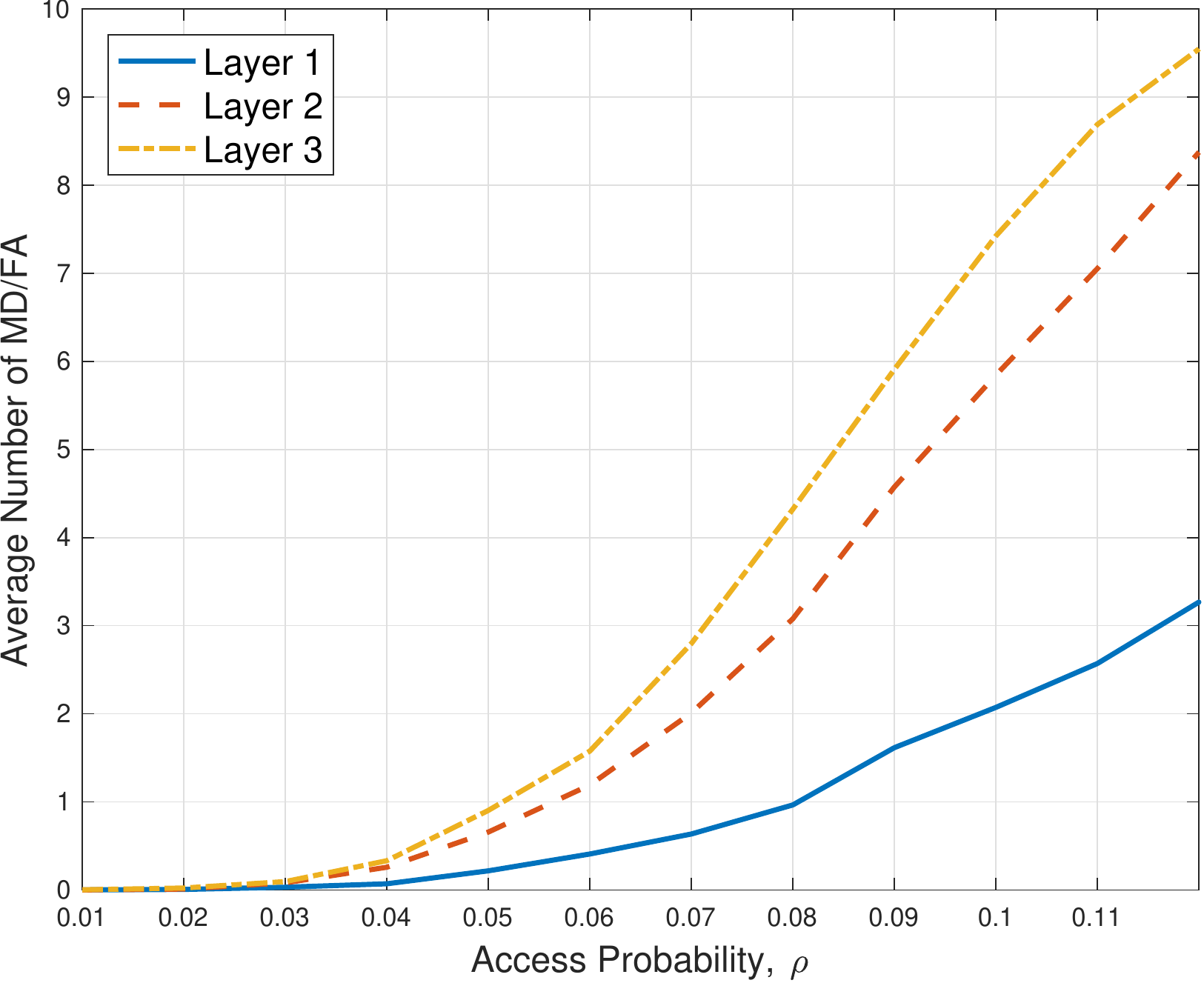} \\
(b) \\
\end{center}
\caption{The average number of MD/FA devices
as a function of access probability
when $K = 300$, $N = 30$, $\Gamma = 4$, $T = 100$, 
and $N_{\rm run} = 5$:
(a) the total number of MD/FA devices with $Q \in \{1,2,3\}$;
(b) the number of MD/FA devices for each layer with $Q = 3$.}
        \label{Fig:plt1}
\end{figure}

We now consider the performance 
of L-CRA depending on the system parameters with
$N_{\rm run} = 5$ for the CAVI algorithm.
In particular, in order to see the advantage of applying NOMA to CRA, 
let us consider the
performances with $Q = 1, 2, 3$ when there are $K = 300$ devices.
Fig.~\ref{Fig:plt1} (a) shows the average total number of MD/FA devices
as a function of the access probability, $\rho = \rho_q$, for all $q$,
when $N = 30$, $\Gamma = 4$, $T = 100$, and $N_{\rm run}
= 5$.
We can see that the average total number of MD/FA devices decreases
with $Q$. When $\rho$ is sufficiently low
(e.g., $\rho < 0.05$), the average total number of MD/FA devices with $Q = 3$
is sufficiently small (e.g., less than 2).
In particular, in the case of $\rho = 0.04$ (where
the average total number of active devices is 12),
the average total numbers of MD/FA devices 
are 6.71, 1.90, and 0.66 when $Q = 1$, 2, and 3, respectively.
Thus, the error rate decreases by a factor of about 3.53 or 10.16 from
conventional CRA (i.e., $Q = 1$) to L-CRA with $Q = 2$ or $3$,
respectively.
However, as mentioned earlier, 
the transmit power at devices increases with $Q$,
which might be the cost of a better performance.

In Fig.~\ref{Fig:plt1} (b),
the average number of MD/FA devices
in each layer is shown  when $Q = 3$.
It is clearly shown that
there is the error propagation as the 
average number of MD/FA devices increases with $q$.
However, although the error propagation exists,
as shown in Fig.~\ref{Fig:plt1} (a), a better performance
can be achieved with a larger $Q$.
As expected, 
the average number of MD/FA devices increases with the access probability,
$\rho$. Note that when $\rho = 0.1$, the average number of 
active devices per layer is 10.
Thus, when $Q = 3$, according to
Fig.~\ref{Fig:plt1} (b), most active devices in layer 3 are incorrectly
detected (i.e., about 8 out of 10).
On the other hand, if $\rho = 0.05$,
the average number of MD/FA devices is approximately 1 in 
layer 3, which means that approximately 20\% of the active devices 
in layer 3 are incorrectly detected.
This implies that the access probability should be low enough 
to avoid excessive re-transmissions.

In the rest of the section, we only consider the case of
$Q = 2$ as the transmit power of the devices in layer 1 
might be too high when $Q = 3$ (which is more
than 24dB) as shown in Table~\ref{TBL:1}.
Note that if $Q = 2$, the 
average transmit power of the devices in layer 1
becomes 16.221dB under the same conditions in Table~\ref{TBL:1}.

\begin{figure}[thb]
\begin{center}
\includegraphics[width=\figwidth]{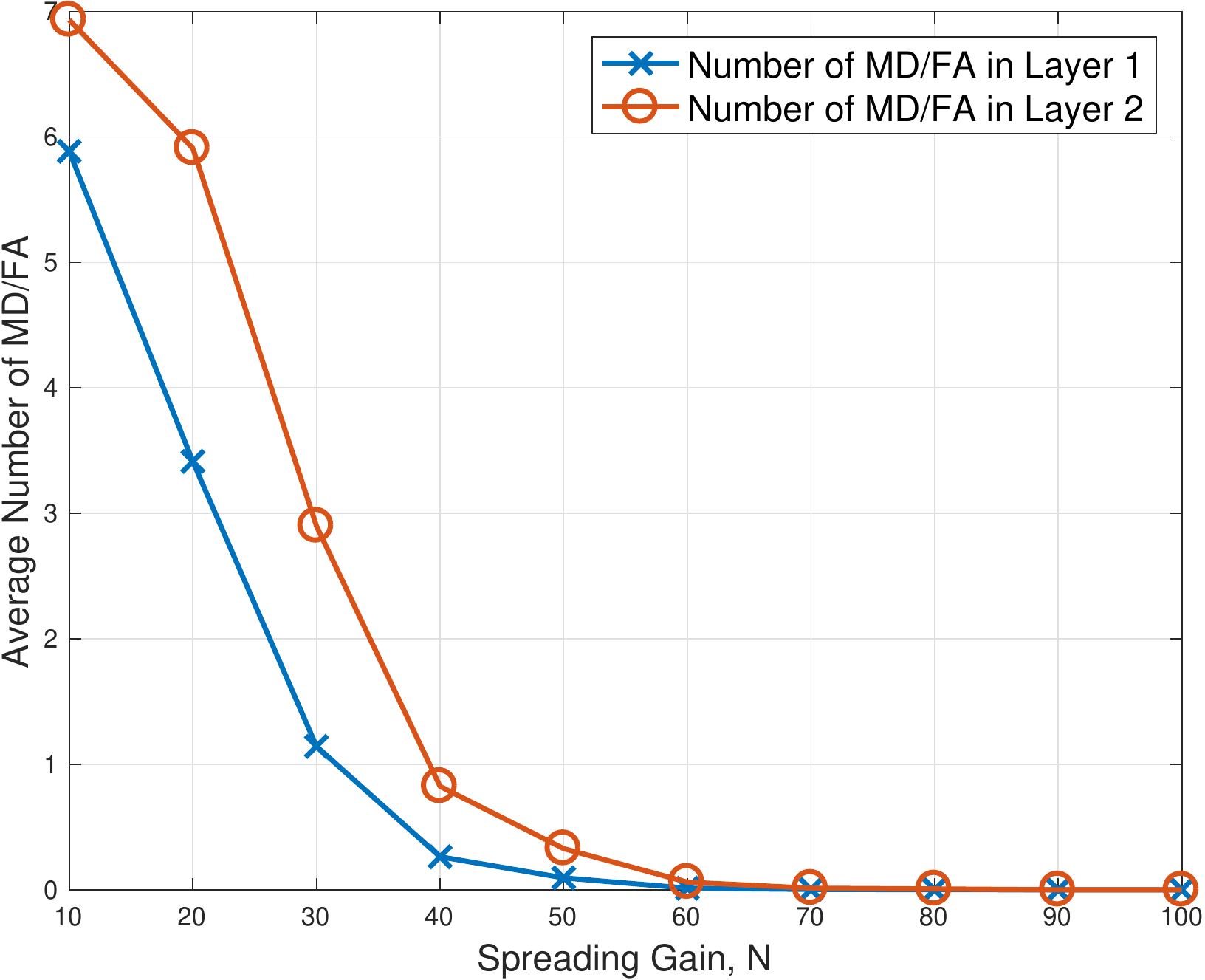} 
\end{center}
\caption{The average number of MD/FA devices
as a function of the spreading gain, $N$,
when $Q = 2$, $M = 150$, $\rho = 0.05$, $\Gamma = 4$, $T = 100$, 
and $N_{\rm run} = 5$.}
        \label{Fig:plt4}
\end{figure}

To see the impact of the spreading gain, $N$,
on the performance, 
the average number of MD/FA devices is shown 
as a function of the spreading gain, $N$,
in Fig.~\ref{Fig:plt4} 
when $Q = 2$, $M = 150$, $\rho = 0.05$, $\Gamma = 4$, $T = 100$, 
and $N_{\rm run} = 5$.
Clearly, the number of MD/FA devices decreases with $N$.
Thus, for a lower probability
of incorrect detection, we need a wider system bandwidth.

\begin{figure}[thb]
\begin{center}
\includegraphics[width=\figwidth]{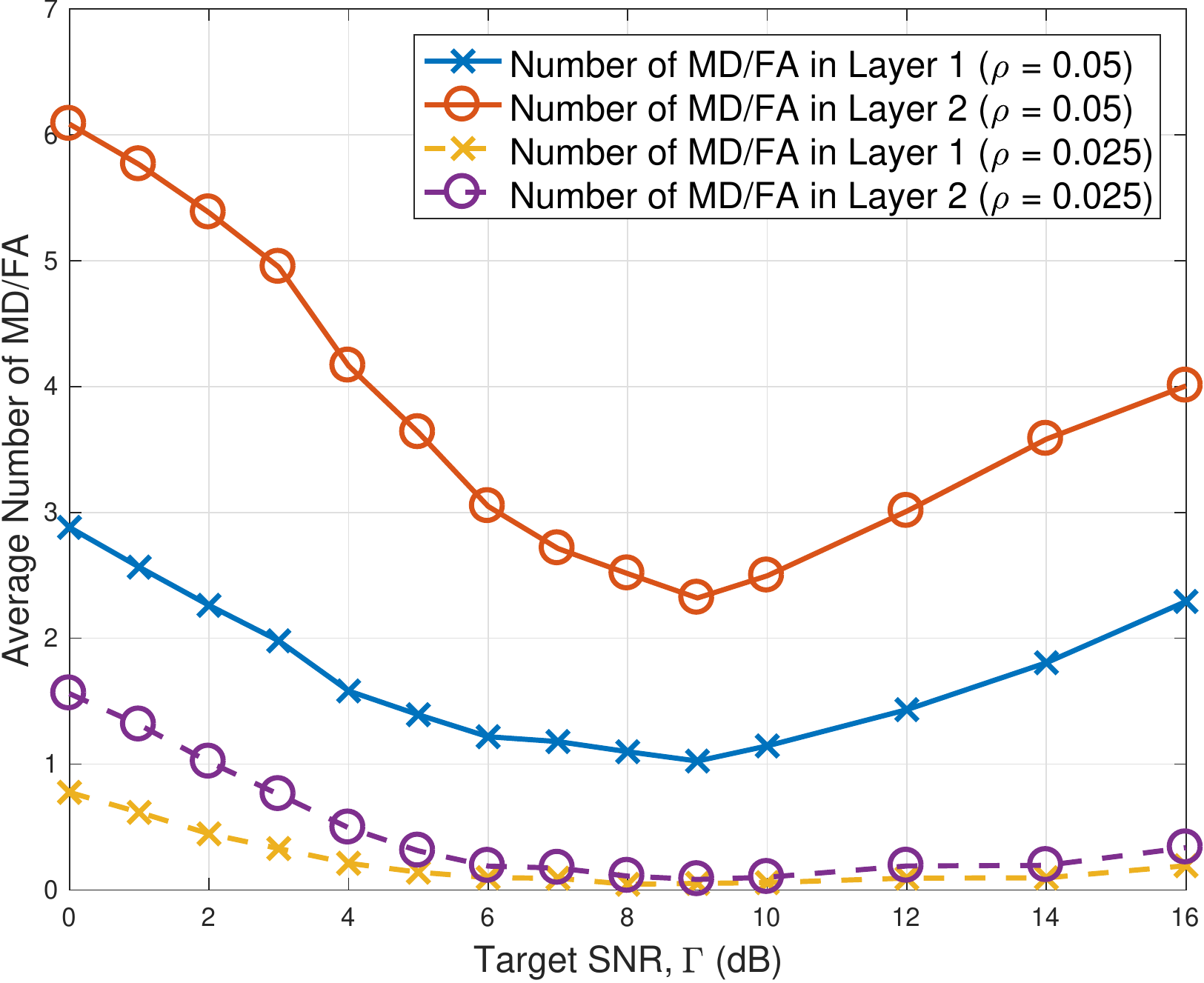} 
\end{center}
\caption{The average number of MD/FA devices
as a function of the target SNR, $\Gamma$,
when $Q = 2$, $M = 150$, $N = 30$, $\rho 
\in \{0.025, 0.05\}$, $T = 100$, 
and $N_{\rm run} = 5$.}
        \label{Fig:plt5}
\end{figure}

Fig.~\ref{Fig:plt5} shows
the average number of MD/FA devices
as a function of the target SNR, $\Gamma$,
when $Q = 2$, $M = 150$, $N = 30$, $\rho 
\in \{0.025, 0.05\}$, $T = 100$, 
and $N_{\rm run}) = 5$.
The average number of MD/FA devices
decreases with $\Gamma$ until $\Gamma$ reaches 9dB. 
However, as $\Gamma$ further increases from 9dB, the 
average number of MD/FA devices increases.
This indicates that a too high target SNR does not help
improve the performance. When the signals in layer 1 are detected,
the superposition of the signals in layer 2 becomes 
the dominant term in the interference-plus-noise, $\bu_{1, (t)}$,
for a high $\Gamma$. As explained in Subsection~\ref{SS:GAI},
the superposition of the signals in layer 2 
is not exactly Gaussian and has a tail probability
that is heavier than that of Gaussian.
This results in a higher error probability. 
Note that at a high target SNR (say $\Gamma = 16$dB),
the performance degradation 
when $\rho = 0.025$ is less severe than that when
$\rho = 0.05$. This is due to the fact that the background noise 
becomes more dominant than the interference with a lower $\rho$,
which leads to a relatively lower tail probability and less
performance degradation.
In general, it is desirable to 
avoid a high target SNR, which is also preferable in terms of
keeping the transmit power of devices low.

\section{Concluding Remarks}	\label{S:Conc}

In this paper, we have applied 
power-domain NOMA to CRA, which results in L-CRA,
to improve the performance of CRA without increasing the system bandwidth.
While the application of NOMA to CRA was straightforward,
the main issue was the determination of the power levels
for multiple layers in the power domain to allow successful
user activity detection with a high probability.
Through a large-system analysis with
Gaussian spreading, we derived design criteria to determine
the power levels. Since 
the MAP detection was used in deriving the design criteria,
a low-complexity detection method that provides approximate MAP solution
was presented using a variational inference approach, namely
the CAVI algorithm.
Simulation results showed that 
the error rate of the device activity detection
can decrease by a factor of about 3.53 or 10.16 from 
conventional CRA 
to L-CRA with two or three layers, respectively,
with a low access probability,
which demonstrates that L-CRA could have a higher
throughput or support more devices
than conventional CRA.

\bibliographystyle{ieeetr}
\bibliography{mtc}

\end{document}